\documentclass[prb,twocolumn,superscriptaddress,showpacs,amsmath,amssymb]{revtex4-1} 
\usepackage{graphicx}
\usepackage{amsmath}
\usepackage{dcolumn}
\usepackage{bm}

\begin{document}

\title{Electronic structure of a Josephson vortex in a SIS junction}

\author{Vadim~Plastovets}
\affiliation{Institute for Physics of Microstructures, Russian
Academy of Sciences, 603950 Nizhny Novgorod, GSP-105, Russia}
\affiliation{University Bordeaux, LOMA UMR-CNRS 5798, F-33405 Talence Cedex, France}
\author{A.~S.~Mel'nikov}
\affiliation{Institute for Physics of Microstructures, Russian
Academy of Sciences, 603950 Nizhny Novgorod, GSP-105, Russia}
\affiliation{Lobachevsky State University of Nizhni Novgorod,
603950 Nizhni Novgorod, Russia}

\date{\today}

\begin{abstract}

The Josephson vortex formed in a superconductor-insulator-superconductor (SIS) junction can affect the quantum mechanics of quasiparticles by creating an effective adiabatic potential determined by the inhomogeneous distribution of the phase difference of the order parameter along the junction. 
Starting from the quasiclassical version of the Bogoliubov-de Gennes (BdG) theory, we found the quasiparticle spectrum and the local density of states (DOS) both for the isolated Josephson vortex and the vortex chain. 
The spatially resolved DOS reveals a peculiar two-peak structure for
each Josephson vortex, which can be detected experimentally using the scanning tunneling microscopy/spectroscopy techniques. 

\end{abstract}

\maketitle

\section{Introduction}

The observation of a rich variety of vortex phases in superconductors and superfluids is known to be one of the convincing manifestations of the quantum coherence in these systems. According to a textbook picture (see, e.g. Ref. \cite{Eshrig}) each vortex has a $2\pi$ circulation of the order parameter phase and carries the magnetic flux quantum $\Phi_0=\pi\hbar c/e$ in the bulk systems. Such vortices affect the local gap function and, thus, perturb the quasiparticle spectrum provoking the formation of the subgap quasiparticle states. 
Experimental detection and study of this subgap spectrum can provide information about the nature of the superconducting state, i.e. about the symmetry and structure of the superconducting gap function
\cite{Bruer2016, Du2015, PhysRevLett.101.166407, Nishimori,PhysRevLett.78.4273, suderow2014imaging, PhysRevLett.75.2754, PhysRevLett.87.277002,RevModPhys.79.353, PhysRevB.99.144514, PhysRevLett.119.237001,volovik2009universe}.
This approach to the probing of the gap structure can be applied for different types of vortex systems including a standard Abrikosov vortex in isotropic superconductors and strongly disturbed vortex solutions in anisotropic or layered superconductors with Josephson interaction between the layers\cite{RevModPhys.66.1125,Brandt_1995, Roditchev2015}.

The electronic structure of a singly quantized Abrikosov vortex has been studied for decades both experimentally and theoretically \cite{RevModPhys.66.1125, Kramer1974,PhysRevB.41.4819, PhysRevLett.64.2711}.
The vortex has a normal core with a radius of the order of the coherence length $\xi=\hbar V_F /\Delta_0$ with Fermi velocity $V_F$ and bulk gap value $\Delta_0$,  surrounded by a circulating supercurrent which reaches a depairing value $j_d$ at the core boundary and decays at the magnetic penetration depth $\lambda$. Circulation of the phase of the order parameter is responsible for the formation of the subgap bound quasiparticle states, which form a so-called anomalous spectral branch, originally discovered in the work of Caroli-de Gennes-Matricon (CdGM)\cite{CAROLI1964307}. 
In the quasiclassical limit $k_F\xi \gg 1$, where $k_F$ is the Fermi momentum, the quasiparticles propagate along the straight classical trajectories, which can be parametrized by the impact parameter $b=-\mu/k_{\bot}$, where $\mu$ is the angular momentum (half an odd integer) defined with respect to the vortex axis and $k_{\bot}$ is the momentum component perpendicular to the vortex axis.
The anomalous spectral branch crosses the Fermi level and varies from $-\Delta_0$ to $\Delta_0$ as $\mu$ changes. The low-energy CdGM spectrum is a linear function of the angular momentum $\mu$: $E_{CdGM}\approx-\mu \Delta_0/k_F\xi$.

In real superconducting crystals, the spectral features described above can be strongly affected by defects of different nature, such as columnar defects, point impurities, and twinning planes.
In particular, these inhomogeneities can modify the shape of the vortex core and, consequently, the subgap spectrum. Another aspect of the influence of the defects on the vortex electronic structure originates from the elastic scattering of quasiparticles at the defect potential profile. 
The consequences of this scattering effect have been investigated for point impurities \cite{PhysRevB.57.5457, Skvortsov1998, PhysRevB.60.14597,PhysRevB.103.024510}, for columnar\cite{PhysRevB.79.134529,PhysRevB.84.134521} and planar \cite{Kawakami_2014,PhysRevB.102.174501,khodaeva2021vortex} defects and for a vortices near the surface of various shape \cite{PhysRevLett.93.247001,PhysRevB.78.064513,PhysRevB.71.214508}.
It has been shown \cite{PhysRevB.102.174501}, in particular, that for a vortex pinned at a high-transparent insulating plane the electron scattering can cause the essential changes in the structure of the low-energy part of the CdGM spectrum even without perturbation of the vortex core shape. This leads to a significant increase in the spectral minigap at the Fermi level and deviation of the low-energy spectrum structure from the above equidistant behavior. 
Recently, this problem has been also addressed in Ref. \cite{khodaeva2021vortex} for the case of vortices shifted from the defect plane or pinned by several intersected linear defects.

The changes in the quasiparticle spectrum of the vortex pinned at a linear defect are of particular interest in the context of the problem of manipulating of the topologically protected Majorana states. 
\cite{PhysRevB.84.075141,PhysRevLett.106.077003,PhysRevB.86.035441,PhysRevB.89.085409,PhysRevLett.100.096407,PhysRevB.82.094522}
The controllable motion of the vortex along the linear defect in an exemplary hybrid structure consisting of a primary superconductor with conventional pairing and a two-dimensional (2D) layer with a nontrivial topology provides a unique possibility to manipulate the Majorana state located in the 2D layer. The value of the minigap in the vortex spectrum in the primary superconductor is of the crucial importance for the topological protection of these operations.

It is important to note here that the solution presented in Ref. \cite{PhysRevB.102.174501} is not self-consistent in the sense that it does not take into account the deformation of the vortex core and the redistribution of the order parameter phase along the defect. 
The self-consistent numerical analysis of the quasiparticle spectrum and DOS for a vortex pinned by the plane defect has been carried out in Ref. \cite{Kawakami_2014} on the basis of the BdG theory for a two-dimensional tight-binding model on a square lattice.
The effect of perturbation correction to the gap profile has been discussed in Ref. \cite{khodaeva2021vortex}.
Such approximation is valid as long as the electronic transparency of the barrier $\mathcal{T}$ is close to unity. 
This limit allows one to observe the changes in the LDOS distribution in the vortex area corresponding to the transition from the pinned Abrikosov vortex to the intermediate Abrikosov-Josephson vortex regime \cite{PhysRevB.46.3187, PhysRevB.75.020504, PhysRevB.77.132502}. However, a generic problem of the electronic structure of the disturbed vortex pinned by a defect with arbitrary transparency remained unsolved.

It is the goal of this paper to suggest a theoretical description of the electronic structure of a vortex pinned by a \textit{low-transparent} defect. The spatial distribution of the order parameter phase in this limit becomes strongly anisotropic, and the corresponding circulating supercurrent along the defect is characterized by the length $\ell$ strongly exceeding $\xi$.
This extreme anisotropy allows one to consider the quasiparticle motion along the junction in the adiabatic approximation. We found that the inhomogeneous distribution of the phase difference or a phase soliton corresponding to the Josephson vortex can form an effective semiclassical potential well for the trapped quasiparticles with subgap energies. The turning points for the quasiparticle motion in this well are responsible for the local increase of the semiclassical wavefunctions providing, thus, a two peak structure in the profile of the local density of states along the junction. 
The distinctive feature of the LDOS pattern for the Josephson vortex under consideration is that the distance between the LDOS peaks is determined by the Josephson length and can well exceed the corresponding distance for Abrikosov vortices. In the high-resolution scanning tunneling microscopy (STM) and scanning tunneling spectroscopy (STS) measurements  these features can be obviously viewed as the spectral signatures of the Josephson vortex.
The paper is organized as follows.
We introduce the basic equations of the BdG theory in Sec. \ref{sec:eqns} and describe the general semiclassical approximation for the BdG equations in the presence of Josephson vortices in Sec. \ref{sec:WKB}.
Sec. \ref{sec:LDOS} is devoted to the calculation of the quasiparticle LDOS in various limits. 
We summarize our results in Sec. \ref{sec:summary}.

\section{Basic equations}\label{sec:eqns}
 
\begin{figure}
\centering
\begin{minipage}[h]{1.0\linewidth}
\includegraphics[width=0.9\textwidth]{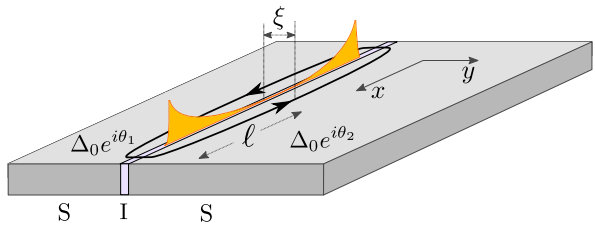}
\end{minipage}
\caption{ {\small Sketch of a SIS structure with a Josephson vortex, created by a circulation of the order parameter phase (black solid line). Spatially separated peaks of the local density of subgap states are schematically shown in orange color.} }
\label{fig1}
\end{figure}

We restrict our consideration to the case of a SIS
system (Fig. \ref{fig1}) in a rather thick superconducting film neglecting all the effects related to the
peculiarities of thin film electrodynamics.
The isolating barrier is positioned at $y=0$ and modeled by the delta function potential $V(y) =\hbar^2k_Fm^{-1}Z_0 \delta(y)$, where $Z_0$ is the dimensionless barrier strength.
The quantum mechanics of quasiparticles in such junction is described by the following BdG equations:
\begin{gather} \label{bdg0}
\begin{pmatrix}
\hat{H}_0-\mu  & \Delta \\
\Delta^*  & \mu-\hat{H}_0
\end{pmatrix}
\hat{\Psi}(x,y)=
E
\hat{\Psi}(x,y),
\end{gather}
where 
$$\hat{H}_0 = -\frac{\hbar^2}{2m}\Big(\frac{\partial^2}{\partial x^2}+\frac{\partial^2}{\partial y^2}\Big)+V(y)$$
is the single-particle Hamiltonian,  $\mu$ is the chemical potential which is equal to the Fermi energy, $\Delta(x,y)$ is the complex-valued gap potential, $\hat{\Psi}(x,y)=(u(x,y),v(x,y))$ is the wave function with electron- $u$ and hole- $v$ like components.
For simplicity we consider here the motion of quasiparticles only in the $x$-$y$ plane assuming that the Fermi surface is a cylinder and therefore neglecting the dependence of the quasiparticle energy on the momentum component $k_z$ along the cylinder axis $z$.
The potential $V(y)$ can be taken into account by introducing specific boundary conditions:
\begin{gather} \label{bc}
\hat{\Psi}(x,+0)=\hat{\Psi}(x,-0) \\ \notag
\frac{\partial \hat{\Psi}}{\partial y}(x,+0)-\frac{\partial \hat{\Psi}}{\partial y}(x,-0)=2k_FZ_0\hat{\Psi}(x,+0)
\end{gather}
For rather low electron transmission through the barrier the system can be described taking a standard approximation for a Josephson junction, i.e., neglecting the spatial dependence of the absolute value of the order parameter $\Delta_0$ and assuming a jumpwise behavior of the superconducting phase:
\begin{gather} \label{Jvort}
\Delta(x,y)=\Delta_0
\begin{cases}
e^{i\theta_2(x)}, \quad y>0 \\ \notag
e^{i\theta_1(x)}, \quad y<0 
\end{cases} 
\end{gather}
The phase difference $\varphi(x)=\theta_1(x)-\theta_2(x)$ is a continuous smooth function changing at a certain length scale $\ell$. 
The length $\ell$ increases with the decrease of the barrier transparency from the values of the order of several superconducting coherence lengths to value of the so-called Josephson penetration depth $\lambda_J = \sqrt{c\Phi_0/16\pi^2 j_c\lambda}$, where $j_c$ is the critical current density through the junction.
The spatial distribution of the function $\varphi(x)$ in the Josephson junction can be obtained from the solution of a standard electrodynamic problem (see e.g. \cite{PhysRevB.46.3187}) which is based on some particular form of the Josephson current-phase relation. The latter, in principle, should be found from the above consideration of the quasiparticle spectrum and wavefunctions. In our further consideration we do not consider the solution of this full self-consistent problem and just analyze the quasiparticle spectral properties for some typical profiles of the superconducting phase. Moreover, in our BdG equations we completely neglect the vector potential assuming, thus, that the supercurrents flowing in superconducting leads are too weak to affect the subgap energy spectrum under consideration.

\section{WKB approximation}\label{sec:WKB}
The model introduced in the previous section contains several important length scales: (i) the Fermi wavelength $k_F^{-1}$; (ii) the typical length scale of the wavefunction decay for the subgap quasiparticles which is roughly the coherence length $\xi$; (iii) the characteristic length of the superconducting phase profile $\ell$.
The Fermi wavelength is certainly the smallest length scale among these values which allows us to use a standard quasiclassical approach, i.e., the so-called Andreev approximation. Moreover, for the junctions with not too large transparency we can introduce an additional simplification valid for the small value of the coherence length compared to the phase distribution length scale $\ell$.

The appearance of the small parameter  $\xi/\ell$ allows one to construct the solution of Eq. ($\ref{bdg0}$) using the semiclassical Wentzel–Kramers–Brillouin (WKB)
approximation. 
Indeed, the slow change of the phase difference $\varphi$ along the junction allows to define the semiclassical energy $E(x,k_x)$ assuming the momentum component $k_x$ and coordinate $x$ to be classical commuting variables. As a next step, we can restore the quantum mechanical commutation rule for these variables using a standard Bohr-Sommerfeld relation. This kind of semiclassical procedure allows to find the true quantum mechanical bound states.

We consider the structure of the wave function in the form $\hat{\Psi}(x,y)=f(x)\hat{g}(y,x)$.
Similarly to the WKB approach, the function $f(x)$ can be written as the following asymptotic expansion:
\begin{gather} \label{f_func}
f(x)=e^{i\kappa^{-1}S(x)}\big[ f_0(x) +\kappa f_1(x)+O(\kappa^2) \big].
\end{gather}
Here $S(x)$ is the eikonal, the functions $S, f_0$ and $f_1$ are real and  $\kappa=(k_F\ell)^{-1}$. After substituting this
solution into the equation ($\ref{bdg0}$) and separating different orders in $\kappa$ we get:
\begin{gather} \label{k_0}
\kappa^0 :  
\Bigg[-\hat{\tau}_3\frac{\hbar^2}{2m}\Big(\frac{\partial^2}{\partial y^2}
+k^2_F-(S')^2 \Big)+ \\ \notag
\begin{pmatrix}
-E  & \Delta(x) \\
\Delta^*(x)  & -E
\end{pmatrix}
\Bigg]\hat{g}f_0=0 
\end{gather}
\begin{gather} \label{k_1}
\kappa^1 : \quad 
-\frac{\hbar^2}{2m}\Big(iS''f_0+2iS'f'_0  \Big)\hat{\tau}_3\hat{g}
= \\ \notag
\Bigg[-\hat{\tau}_3\frac{\hbar^2}{2m}\Big(\frac{\partial^2}{\partial y^2}
+k^2_F-(S')^2 \Big)+
\begin{pmatrix}
-E  & \Delta(x) \\
\Delta^*(x)  & -E
\end{pmatrix}
\Bigg]\hat{g}f_1,
\end{gather}
where the prime means the derivative $\partial/\partial x$ and $\hat{\tau}_3$ is the Pauli matrix in the electron-hole Nambu space.
The equation ($\ref{k_0}$) contains a one-dimensional equation
which together with the boundary conditions ($\ref{bc}$) can be viewed as the short SIS junction problem \cite{PhysRevLett.66.3056, PhysRevLett.67.3836} in which, due to the semiclassical approximation, the momentum $k_x$ is replaced by $S'$, and the coordinate $x$ is a parameter.
Introducing an auxiliary equation 
\begin{align}\label{bdg_y}
-\hat{\tau}_3\frac{\hbar^2}{2m}\Big(\frac{\partial^2}{\partial y^2}
+k^2_F-(S')^2 \Big)\hat{g}+
\begin{pmatrix}
0  & \Delta(x) \\
\Delta^*(x)  & 0
\end{pmatrix}
\hat{g}=\omega(x) \hat{g}
\end{align}
and using ($\ref{bc}$) we obtain a quasiparticle spectrum $\omega$ in the presence of a ``frozen" phase distribution $\varphi(x)$.
Eigenfunctions $\hat{g}(y,x)=(g_u,g_v)^T$ can be written as:
\begin{align} \label{gfunc}
    \hat{g}(y,x)=e^{i\text{sign}(y)\frac{\varphi(x)}{4}\hat{\tau}_3}
    \left\{
    \begin{array}{ll}
        \Big(c_{1}e^{- iq_+y}+d_{1}e^{ iq_-y}\hat{\tau}_1\Big) \hat{g}_0 & y<0 \\
        \\
        \Big(c_{2}e^{ iq_+y}+d_{2}e^{- iq_-y}\hat{\tau}_1\Big)  \hat{g}_0 & y>0
    \end{array}
    \right.
\end{align}
Here the vector $\hat{g}_0=(\tilde{u},\tilde{v})^T$ has the electron-like and hole-like parts:
\begin{gather} \notag
\tilde{u}=\frac{1}{\sqrt{2}}\sqrt{1+i\frac{\sqrt{\Delta_0^2-\omega^2}}{\omega}} , \quad
\tilde{v}=\frac{1}{\sqrt{2}}\sqrt{1-i\frac{\sqrt{\Delta_0^2-\omega^2}}{\omega}}
\end{gather}
and the wave vector is
\begin{gather} \notag
q_{\pm}=\sqrt{k^2_F-(S')^2}\pm i \frac{m}{\hbar^2}\frac{\sqrt{\Delta_0^2-\omega^2}}{\sqrt{k^2_F-(S')^2}}.
\end{gather}
Note, that in the latter expression we use the expansion in the parameter $(k_F\xi)^{-1}$, which is valid due to the quasiclassical condition $\Delta_0/E_F\sim (k_F\xi)^{-1} \ll 1$.
The coefficients $c_{1,2}$ and $d_{1,2}$ are determined by the boundary conditions ($\ref{bc}$) and normalization condition:
$$
\int  \hat{g}^{\dagger }(y,x)\hat{g}(y,x)dy=k_F^{-1}.
$$
With the help of equations ($\ref{gfunc}$) and ($\ref{bc}$) we obtain the resulting spectrum of such a system, which is essentially the spectrum of the short SIS system \cite{PhysRevLett.67.3836}
\begin{gather} \label{spec}
\omega(x)=\pm\Delta_0\sqrt{1-\mathcal{T}\sin^2\Big( \frac{\varphi(x)}{2} \Big)},
\end{gather}
where 
\begin{gather} \notag
\mathcal{T}=
\Bigg(1+\frac{k_F^2Z^2_0}{k^2_F-(S'(x))^2}\Bigg)^{-1}.
\end{gather}
Substituting (\ref{bdg_y}) into (\ref{k_0}) and using the condition
of existence of a nontrivial solution we easily get 
\begin{gather} \label{det}
E=\omega(x).
\end{gather}

To obtain $f_0$ from the  matrix equation (\ref{k_1}) one has to use the Fredholm theorem\cite{Fredholm} which gives us the solvability condition for the Eq. (\ref{k_1}):
\begin{gather} \notag
f_0\hat{g}^{\dagger} \Big( -\frac{\hbar^2}{2m}i\big[ S''\hat{f}_0+2S'\hat{f}'_0 \big]\hat{\tau}_3\hat{g} \Big)=0, 
\end{gather}
which can be rewritten as
\begin{gather} \label{Fr2}
\big( \hat{g}^{\dagger}\hat{\tau}_3\hat{g} \big) \frac{\partial}{\partial x}\big( f^2_0 S' \big)=0, 
\end{gather}
The term $\big( \hat{g}^{\dagger}\hat{\tau}_3\hat{g} \big)$
can be expressed as
$$
\big( \hat{g}^{\dagger}\hat{\tau}_3\hat{g} \big) \sim \frac{1}{S'}\frac{\partial \omega}{\partial S'}
$$
using equation (\ref{bdg_y}) 
and it tends to zero only at the specific points $\varphi(x)=2\pi k$ with an integer k.
Therefore, we come to the equation $\big( f^2_0 S' \big)'=0$ from which the function $f_0$
can be found. In the first-order WKB approximation, the function (\ref{f_func}) has a standard form:
\begin{gather} \label{wf}
f(x)=\frac{C}{\sqrt{S'}} e^{\pm i\int S'(\xi)d\xi },
\end{gather}
where $C$ is an arbitrary constant. The quantity $S'$ is the gradient 
of the eikonal, therefore it can be interpreted as the x-component of the classical local quasiparticle momentum  and here it is convenient to use the notation $S'\equiv k_x$. The expression for this momentum follows from ($\ref{det}$):
\begin{gather} \label{Sx}
\frac{k_x(x)}{k_F} =\sqrt{1+Z^2_0\frac{\Delta_0^2-E^2}{\Delta_0^2\cos^2(\varphi(x)/2)-E^2}}.
\end{gather}

\begin{figure} 
\centering
\begin{minipage}[h]{1.0\linewidth}
\includegraphics[width=1.0\textwidth]{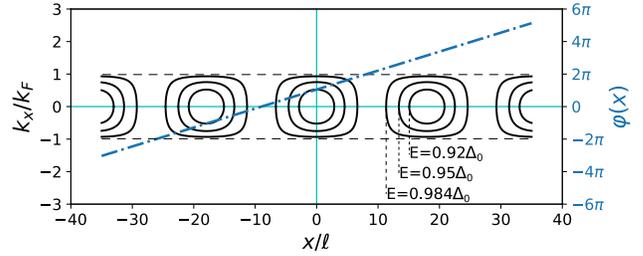}
\end{minipage}
\caption{ {\small Semiclassical orbits (black lines) in the plane $(k_x,x)$ described by the Eq. (\ref{Sx}) for the linearly growing phase difference $\varphi(x)$ (blue dash-dotted line) for a fixed value of the parameter $Z_0=2$ and different energies $E$. Each set of concentric orbits corresponds to a single vortex in a vortex chain.}}
\label{fig1_1}
\end{figure}

The above dependence of the momentum $k_x$ on the coordinate $x$ at a fixed energy $E$ allows one to view the motion of quasiparticles in the presence of the superconducting phase profile as the motion in a smooth adiabatic potential. This potential has a set of turning points where $k_x(x)=0$, therefore one can define closed semiclassical orbits in the plane $(k_x, x)$. A set of exemplary semiclassical orbits for the particular case of a linearly growing phase difference $\varphi(x)=0.35(x/\ell)+\pi$ is shown in Fig. \ref{fig1_1}. 
Obviously, to restore the true quantum mechanics one can apply the Bohr-Sommerfeld quantization rule
\begin{gather} \label{BZ}
\oint k_x(x)dx=2\pi (n+\beta),
\end{gather}
where $n$ is an integer, and obtain the discrete spectrum levels. Certainly, the interlevel distance should be small due to the small parameter $\kappa$.  Similarly to a standard quasiclassical version of the quantum mechanics the parameter $\beta$ is of the order of the unity and its effect on the behavior of the discrete energy spectrum at large quantum numbers $n$ is rather weak. In order to determine the appropriate value of $\beta$ one needs to solve the quantum mechanical problem near the turning points beyond the quasiclassical approximation. This calculation is beyond the scope of our work.

Note, that the value of $k_x$ does not exceed $k_F$ in the classically allowed region, but the denominator of Eq. (\ref{Sx}) tends to zero at some singular points. This is a direct consequence of using of the "frozen" phase approximation; however, since these points are in the forbidden region, the semiclassical approximation is not violated.

It should be noticed that the lower bound of the spectrum is determined by the general expression $E_b=\Delta_0Z_0 / \sqrt{1+Z^2_0}$ for a minimal energy value of localized states $\omega(x)$. For rather large values $Z_0$ this condition means that all the features related to the bound states in the Josephson junction can be observed only at energies rather close to $\Delta_0$. The expression for the low-lying energy levels close to $E_b$ can be obtained explicitly from the Eq. (\ref{BZ}). Using a linearized expression for the phase difference $\varphi\approx ax+\pi$ with a slope $a$ in the vicinity of the orbit center and assuming the condition $Z_0\gtrsim 1$ in the Eq. (\ref{Sx}) we find the discrete spectrum
\begin{gather} \label{spectrum}
E_n\approx E_b\sqrt{1+\frac{(n+\beta)a\pi\kappa}{Z_0\sqrt{1+Z_0^2}}}.
\end{gather}
It is interesting to note that the square root dependence of the spectrum on the level number $n$ has already been observed in the case of an Abrikosov vortex pinned at a high-transparent defect with $Z_0\ll 1$ \cite{PhysRevB.102.174501}. 
In such a system the deformation of the bound CdGM states in a vortex core results in the appearance of a "hard" minigap, which determines the value of the lowest energy level in the spectrum, besides this the electron scattering
at the defect plane also provides a "soft" minigap which is $\Delta_{soft}\sim\Delta_0Z_0$. Although direct comparison of (\ref{spectrum}) and the result for the high-transparency limit is not possible, it can be seen that quantitatively this "soft" minigap $\Delta_{soft}$ coincides with the lowest energy level $E_{n=0}\approx\Delta_0Z_0$ from (\ref{spectrum}), which is actually a "hard" minigap in the present system where the vortex core is absent. 
The energy value $\Delta_0Z_0$ always appears in the systems with a barrier of the finite transparency; therefore, one can expect this quantity to play an essential role throughout the entire crossover from the pinned Abrikosov to the Josephson vortex with an increase of the barrier strength $Z_0$.

Finally, we get the adiabatic solution of the BdG problem $\hat{\Psi}(x,y)$, which consists of two parts:
$\hat{g}$ from (\ref{gfunc}) and $f$ from (\ref{wf}). Following the standard procedure of constructing semiclassical solution in the potential well \cite{Landau} we find the function $f(x)$ which has an oscillating behavior in the classically allowed region and decays exponentially in the classically forbidden region.  

\section{Local density of states}\label{sec:LDOS}

As we discussed in the previous section, the distance between the true quantum mechanical levels appears to be extremely small due to the small value of the inverse quasiclassical parameter $\kappa$. For example, a low-lying part of the discrete spectrum (\ref{spectrum}) provides $E_{n+1}-E_n\approx a\pi\kappa\Delta_0/2(1+Z^2_0)$.
Considering possible experimentally measurable hallmarks of the subgap quasiparticle states it may be much more important to analyze the local density of states in the semiclassical limit neglecting the level quantization. An appropriate expression for the local DOS reads:
\begin{gather} 
\nu(x,y,E)
= k_F\int^{k_F}_{-k_F} \frac{dk_x}{2\pi}|g_u(x,y)|^2
\delta\big(E-\omega(k_x)\big),
\end{gather} 
where the function $g_u(x,y)$ is defined in (\ref{gfunc}). 
Evaluating the integral we find:
\begin{gather} \label{ldos}
\frac{\nu(x,y, E)}{\nu_{2D}}=\frac{|g_u|^2}{\tilde{k}_x}
\frac{2E/\Delta_0\big(1-\tilde{k}_x^2\big)\big(1-\tilde{k}_x^2+Z^2_0\big)}{Z^2_0(1-E^2/\Delta^2_0)},
\end{gather} 
where $\nu_{2D}=m/\pi\hbar^2$ is a local density of states of a two-dimensional electron gas and dimensionless momentum $\tilde{k}_x(x)=k_x/k_F$ is taken from (\ref{Sx}).
A singularity $\nu(k_x\rightarrow 0)\sim k_x^{-1}$ in the vicinity of each turning point $k_x=0$ should be both regularized by a more accurate solution of a WKB problem and smeared by various broadening effects.
Since the position of these peculiarities is defined by the turning points, their existence is restricted by the energy interval $E_b<E<\Delta_0$, as it was discussed above. 

\subsection{Single Josephson vortex}

\begin{figure}[] 
\centering
\begin{minipage}[h]{1.0\linewidth}
\includegraphics[width=0.95\textwidth]{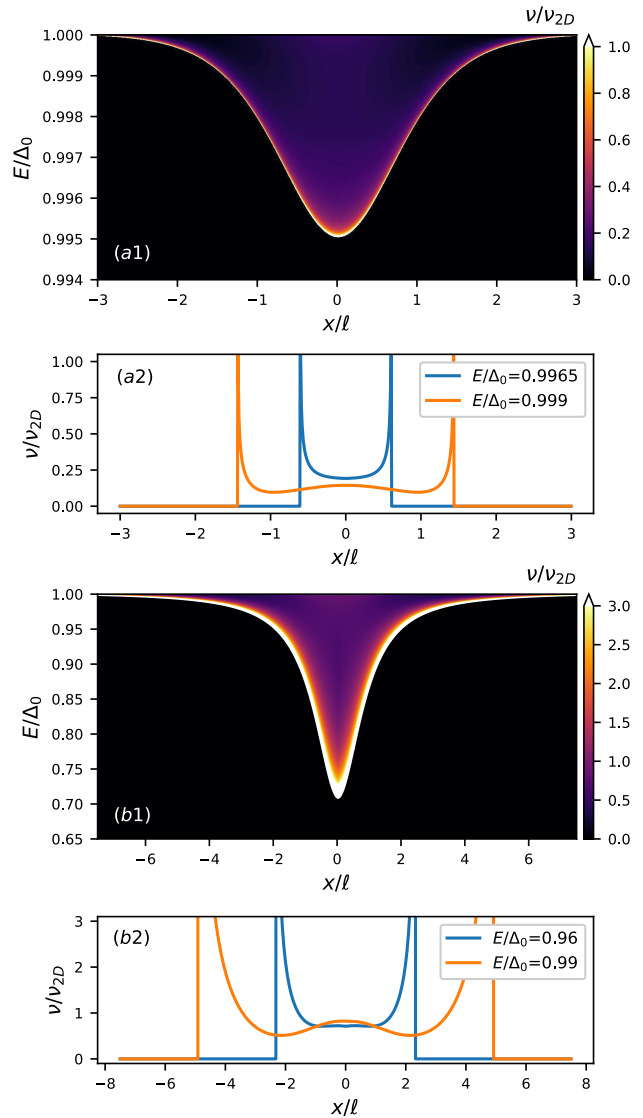}
\end{minipage}
\caption{ {\small 
Distribution of the LDOS of the quasiparticles in the Josephson vortex 
$\nu(x,y=0, E)$ as a function of the energy $E$ and coordinate along the barrier $x$ for local $(a1)$ (\ref{FP}) and nonlocal $(b1)$ (\ref{Nonlocal}) regimes. Subplots $(a2)$ and $(b2)$ show the cross section of the LDOS at a certain energy value for $Z_0=10$ (a) and $Z_0=1$ (b). 
The maximum values are truncated at $\nu/\nu_{2D}=(1,3)$ for (a,b) for illustrative purposes.
 }}
\label{fig2}
\end{figure}

Now we proceed with consideration of several specific models for the phase distribution $\varphi(x)$. 
First, consider the limit $j_c\ll j_d\xi/\lambda$, which is realized for a low transparent insulating barrier with $Z_0\gg \sqrt{12\pi^2\lambda/\xi}$ \cite{PhysRevB.102.174501,PhysRevB.46.3187}. In this case the electrodynamics of the Josephson system is local therefore the phase distribution obeys the sine-Gordon equation with the well-known soliton solution \cite{Tinkham04}:
\begin{gather} \label{FP}
\tilde{\varphi}(x)=4\arctan e^{x/\ell},
\end{gather}
which corresponds to a single isolated Josephson vortex with the size of $\ell\sim\lambda_J\gg\lambda\gg\xi\gg k^{-1}_F$. Two last inequalities assume the limit of a strong type-II superconductor and the validity of the quasiclassical approximation described above. 
With the help of the relation $\sin^2(\tilde{\varphi}(z)/2)=\cosh^{-2}(z)$ we obtain an explicit expression for the turning points:
\begin{gather} \label{tpoints}
\frac{x_{a,b}}{\ell}=\pm \ln\Bigg(\frac{\Delta_0-\sqrt{\Delta_0^2-(\Delta_0^2-E^2)(1+Z^2_0)}}{\sqrt{(\Delta_0^2-E^2)(1+Z^2_0)}}\Bigg).
\end{gather}
Using the expression (\ref{Sx}) for $k_x(x)$ and the wave function $\hat{g}$  we can plot the dependence of the LDOS (\ref{ldos}) on the coordinate along the junction $x$ directly at the junction line $y=0$. 
A typical example of the spatial distribution of LDOS for different energy values is shown in Fig. \ref{fig2}(a). 
The local DOS along the junction clearly reveals two peaks (schematically shown in Fig. \Ref{fig1}).  
The formation of these peaks, which are essentially signatures of the Josephson vortex, is a direct consequence of the semiclassical motion of trapped quasiparticles described above.
At the same time, the exact form of the function $\varphi(x)$ does not qualitatively affect the formation of closed orbits in the plane $(k_x, x)$. Therefore, the observation of the above spectral features is possible for various kinds of $2\pi$-soliton, proposed for different parameters of the Josephson SIS junction \cite{PhysRevB.46.3187}. 

For example, one can consider a so-called nonlocal regime of a Josephson junction, which is realized for the opposite limit $j_d \gg j_c\gg j_d\xi/\lambda$. In our model, this limit can be realized when the transparency of the barrier is sufficiently low, i.e. $1 \lesssim Z_0\lesssim\sqrt{12\pi^2\lambda/\xi}$. In such a case the nonlocal equation for the phase has a soliton-like solution 
\begin{gather} \label{Nonlocal}
\tilde{\varphi}(x)=\pi+2\arctan(x/\ell),
\end{gather}
which corresponds to a single Josephson-Abrikosov vortex with the size of $\ell$, where $\lambda\gg\lambda_J \gg \ell\gg \xi$. As in the local case, this solution assumes the condition $\kappa=(k_F\ell)^{-1}\ll1$ to be fulfilled, therefore, it is possible to use the WKB approximation for (\ref{Nonlocal}).
With the help of relation $\sin^2(\tilde{\varphi}(z)/2)=(1+z^2)^{-1}$ we obtain an explicit expression for the turning points:
\begin{gather} \label{NL_tpoints}
\frac{x_{a,b}}{\ell}=\pm \sqrt{\frac{\Delta_0^2}{(\Delta_0^2-E^2)(1+Z^2_0)}-1}.
\end{gather}
Spatial dependence of quasiparticle LDOS for a nonlocal vortex is shown in Fig. \ref{fig2}(b). 
Considering both limits we find that described peculiarities in the LDOS can be observed in the wide range of transparencies.

Let us note that some basic features of the LDOS and quasiparticle spectrum discussed above are qualitatively close to the ones predicted in Ref. \cite{Kawakami_2014} on the basis of the numerical BdG calculations. This qualitative similarity reveals itself, in particular, in the behavior of the lowest energy level of the subgap spectrum: the energy of this level grows with the decreasing barrier transparency (i.e., the hopping strength at the defect line) resulting in the suppression of the LDOS at low energies and the splitting of the subgap energy peak (see Figs. 2 and 3 of the Ref. \cite{Kawakami_2014}). The quantitative comparison is however difficult since our calculations are based on the quasiclassical approach assuming rather large ratio $E_F/\Delta_0$, while in the Ref. \cite{Kawakami_2014} this ratio is not so large.

\subsection{Array of Josephson vortices}

\begin{figure}[h!] 
\centering
\begin{minipage}[h]{1.0\linewidth}
\includegraphics[width=1.0\textwidth]{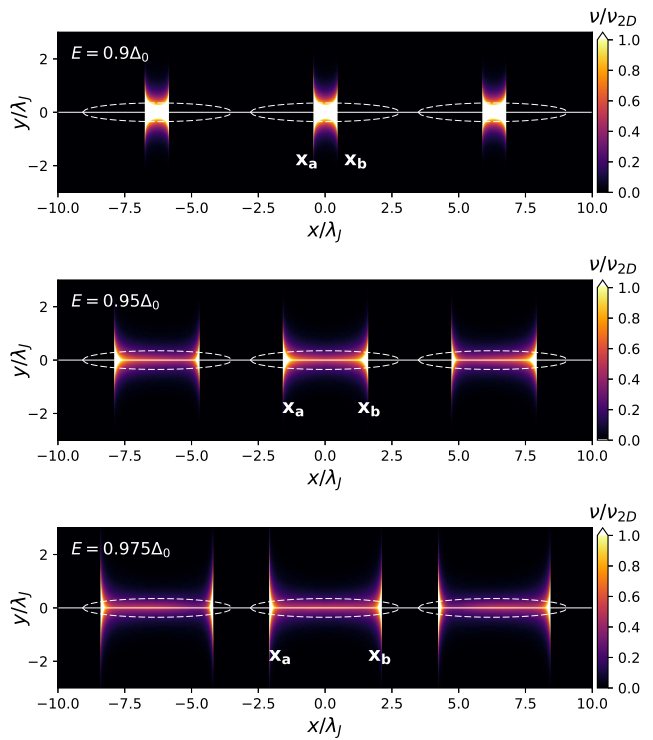}
\end{minipage}
\caption{ {\small Spatial distribution of the quasiparticle LDOS $\nu(x, y, E)$ for a set of three Josephson vortices, schematically shown by the dashed lines for the different values of the energy. White solid line shows the barrier position. The positions of the peaks of LDOS $x_{a,b}$ correspond to the turning points (\ref{tp2}) for each vortex. The parameters are: $a=1$, $Z_0=2$, $k_F\xi=30$, $k_F\lambda_J=100$. }}
\label{fig3}
\end{figure}

The idea of formation of an adiabatic potential for quasiparticles in the Josephson vortex holds for a quite general form of the function $\varphi(x)$. 
Consider as an example a general solution of the Ferrell-Prange equation describing the local limit of the Josephson junction
$$
\frac{\lambda_J}{\sqrt{2}}\int^{\varphi}_{\varphi_0}\frac{d\varphi}{\sqrt{C-\cos\varphi}}=x,
$$
where $C$ is a constant, $\varphi_0$ is a phase difference value at $x=0$ and the scale of the spatial distribution of the phase along the junction is $\ell\sim\lambda_J$. The case $C=1$ corresponds to the  phase soliton described in the previous subsection, while at $C>1$ the phase grows continuously and each increase of the phase by $2\pi$ corresponds to a Josephson vortex. For such a solution, an array of semiclassical potential wells is formed and, consequently, we get an array of LDOS peaks corresponding to these wells.

For illustration we take the limit of high magnetic fields and dense vortex lattices which corresponds to the values $C\gg 1$. The solution can be chosen in the following form: $\tilde{\varphi}(x)=ax/\lambda_J+\pi$, where the constant $a$ is proportional to the external magnetic field in the contact. Then, using (\ref{Sx}) we find a set of the turning points
\begin{gather} \label{tp2}
\frac{x_{a,b}}{\lambda_J}=\mp \frac{2}{a}\arccos\sqrt{(Z^2_0+1)(1-E^2/\Delta^2_0)} \pm \frac{2\pi}{a}n,
\end{gather}
where $n$ is an integer corresponding to different vortices in the vortex array. 
The result is a double-period peak structure shown in Fig. (\ref{fig3}). As the quasiparticle energy approaches the gap value, the distance between the peaks in each vortex increases and the peaks from different vortices approach each other. 
This leads to the coupling of states in the neighboring classically allowed regions, which is not taken into account in the present work.

\section{Summary}\label{sec:summary}

To summarize, we analyzed the subgap spectrum of localized quasiparticle states in a SIS junction with a finite transparency in the presence of an inhomogeneous phase difference along the junction, which corresponds to an array of Josephson vortices. 
Since the spatial scale of the Josephson vortex is usually much larger than the characteristic quasiparticle wavelength, the phase difference profile can be treated as an effective adiabatic potential.  
This potential affects the quasiparticle motion along the barrier and leads to the appearance of the closed semiclassical orbits in the plane of $(S',x)$. We restored the quantum spectrum corresponding to this orbits by using the Bohr-Sommerfeld quantization rule. The obtained discrete spectrum $E_n$ reveals a minigap which increases with an increase in the barrier strength $Z_0$.

We found, that the semiclassical orbits are responsible for the formation of a set of turning points at which momentum along the barrier plane $S'$ goes to zero. Corresponding local increase in the quasiparticle wave function near each turning point leads to the formation of the corresponding peak of the LDOS.
We claim that these peaks can be observed in the STS/STM experiments in both local and nonlocal Josephson junction regimes in a fairly large energy range below the gap.
\acknowledgements
We are grateful to  A. V. Samokhvalov for helpful discussions. The work has been supported by the Russian Science Foundation Grant No. 21-72-10161.


%


\begin{thebibliography}{45}%
\makeatletter
\providecommand \@ifxundefined [1]{%
 \@ifx{#1\undefined}
}%
\providecommand \@ifnum [1]{%
 \ifnum #1\expandafter \@firstoftwo
 \else \expandafter \@secondoftwo
 \fi
}%
\providecommand \@ifx [1]{%
 \ifx #1\expandafter \@firstoftwo
 \else \expandafter \@secondoftwo
 \fi
}%
\providecommand \natexlab [1]{#1}%
\providecommand \enquote  [1]{``#1''}%
\providecommand \bibnamefont  [1]{#1}%
\providecommand \bibfnamefont [1]{#1}%
\providecommand \citenamefont [1]{#1}%
\providecommand \href@noop [0]{\@secondoftwo}%
\providecommand \href [0]{\begingroup \@sanitize@url \@href}%
\providecommand \@href[1]{\@@startlink{#1}\@@href}%
\providecommand \@@href[1]{\endgroup#1\@@endlink}%
\providecommand \@sanitize@url [0]{\catcode `\\12\catcode `\$12\catcode
  `\&12\catcode `\#12\catcode `\^12\catcode `\_12\catcode `\%12\relax}%
\providecommand \@@startlink[1]{}%
\providecommand \@@endlink[0]{}%
\providecommand \url  [0]{\begingroup\@sanitize@url \@url }%
\providecommand \@url [1]{\endgroup\@href {#1}{\urlprefix }}%
\providecommand \urlprefix  [0]{URL }%
\providecommand \Eprint [0]{\href }%
\providecommand \doibase [0]{https://doi.org/}%
\providecommand \selectlanguage [0]{\@gobble}%
\providecommand \bibinfo  [0]{\@secondoftwo}%
\providecommand \bibfield  [0]{\@secondoftwo}%
\providecommand \translation [1]{[#1]}%
\providecommand \BibitemOpen [0]{}%
\providecommand \bibitemStop [0]{}%
\providecommand \bibitemNoStop [0]{.\EOS\space}%
\providecommand \EOS [0]{\spacefactor3000\relax}%
\providecommand \BibitemShut  [1]{\csname bibitem#1\endcsname}%
\let\auto@bib@innerbib\@empty
\bibitem [{\citenamefont {Huebener}\ \emph {et~al.}(2002)\citenamefont
  {Huebener}, \citenamefont {Schopohl},\ and\ \citenamefont
  {Volovik}}]{Eshrig}%
  \BibitemOpen
  \bibfield  {author} {\bibinfo {author} {\bibfnamefont {R.~P.}\ \bibnamefont
  {Huebener}}, \bibinfo {author} {\bibfnamefont {N.}~\bibnamefont {Schopohl}},\
  and\ \bibinfo {author} {\bibfnamefont {G.~E.}\ \bibnamefont {Volovik}},\
  }\href@noop {} {\emph {\bibinfo {title} {Vortices in Unconventional
  Superconductors and Superfluids, Springer Series in Solid-State Sciences Vol.
  132}}}\ (\bibinfo  {publisher} {Springer},\ \bibinfo {year}
  {2002})\BibitemShut {NoStop}%
\bibitem [{\citenamefont {Bru{\'e}r}\ \emph {et~al.}(2016)\citenamefont
  {Bru{\'e}r}, \citenamefont {Maggio-Aprile}, \citenamefont {Jenkins},
  \citenamefont {Risti{\'{c}}}, \citenamefont {Erb}, \citenamefont {Berthod},
  \citenamefont {Fischer},\ and\ \citenamefont {Renner}}]{Bruer2016}%
  \BibitemOpen
  \bibfield  {author} {\bibinfo {author} {\bibfnamefont {J.}~\bibnamefont
  {Bru{\'e}r}}, \bibinfo {author} {\bibfnamefont {I.}~\bibnamefont
  {Maggio-Aprile}}, \bibinfo {author} {\bibfnamefont {N.}~\bibnamefont
  {Jenkins}}, \bibinfo {author} {\bibfnamefont {Z.}~\bibnamefont
  {Risti{\'{c}}}}, \bibinfo {author} {\bibfnamefont {A.}~\bibnamefont {Erb}},
  \bibinfo {author} {\bibfnamefont {C.}~\bibnamefont {Berthod}}, \bibinfo
  {author} {\bibfnamefont {{\O}.}~\bibnamefont {Fischer}},\ and\ \bibinfo
  {author} {\bibfnamefont {C.}~\bibnamefont {Renner}},\ }\href@noop {}
  {\bibfield  {journal} {\bibinfo  {journal} {Nature Communications}\ }\textbf
  {\bibinfo {volume} {7}},\ \bibinfo {pages} {11139} (\bibinfo {year}
  {2016})}\BibitemShut {NoStop}%
\bibitem [{\citenamefont {Du}\ \emph {et~al.}(2015)\citenamefont {Du},
  \citenamefont {Fang}, \citenamefont {Wang}, \citenamefont {Li}, \citenamefont
  {Du}, \citenamefont {Yang}, \citenamefont {Zhu},\ and\ \citenamefont
  {Wen}}]{Du2015}%
  \BibitemOpen
  \bibfield  {author} {\bibinfo {author} {\bibfnamefont {Z.}~\bibnamefont
  {Du}}, \bibinfo {author} {\bibfnamefont {D.}~\bibnamefont {Fang}}, \bibinfo
  {author} {\bibfnamefont {Z.}~\bibnamefont {Wang}}, \bibinfo {author}
  {\bibfnamefont {Y.}~\bibnamefont {Li}}, \bibinfo {author} {\bibfnamefont
  {G.}~\bibnamefont {Du}}, \bibinfo {author} {\bibfnamefont {H.}~\bibnamefont
  {Yang}}, \bibinfo {author} {\bibfnamefont {X.}~\bibnamefont {Zhu}},\ and\
  \bibinfo {author} {\bibfnamefont {H.-H.}\ \bibnamefont {Wen}},\ }\href@noop
  {} {\bibfield  {journal} {\bibinfo  {journal} {Scientific Reports}\ }\textbf
  {\bibinfo {volume} {5}},\ \bibinfo {pages} {9408} (\bibinfo {year}
  {2015})}\BibitemShut {NoStop}%
\bibitem [{\citenamefont {Guillam\'on}\ \emph {et~al.}(2008)\citenamefont
  {Guillam\'on}, \citenamefont {Suderow}, \citenamefont {Vieira}, \citenamefont
  {Cario}, \citenamefont {Diener},\ and\ \citenamefont
  {Rodi\`ere}}]{PhysRevLett.101.166407}%
  \BibitemOpen
  \bibfield  {author} {\bibinfo {author} {\bibfnamefont {I.}~\bibnamefont
  {Guillam\'on}}, \bibinfo {author} {\bibfnamefont {H.}~\bibnamefont
  {Suderow}}, \bibinfo {author} {\bibfnamefont {S.}~\bibnamefont {Vieira}},
  \bibinfo {author} {\bibfnamefont {L.}~\bibnamefont {Cario}}, \bibinfo
  {author} {\bibfnamefont {P.}~\bibnamefont {Diener}},\ and\ \bibinfo {author}
  {\bibfnamefont {P.}~\bibnamefont {Rodi\`ere}},\ }\href@noop {} {\bibfield
  {journal} {\bibinfo  {journal} {Phys. Rev. Lett.}\ }\textbf {\bibinfo
  {volume} {101}},\ \bibinfo {pages} {166407} (\bibinfo {year}
  {2008})}\BibitemShut {NoStop}%
\bibitem [{\citenamefont {Nishimori}\ \emph {et~al.}(2004)\citenamefont
  {Nishimori}, \citenamefont {Uchiyama}, \citenamefont {Kaneko}, \citenamefont
  {Tokura}, \citenamefont {Takeya}, \citenamefont {Hirata},\ and\ \citenamefont
  {Nishida}}]{Nishimori}%
  \BibitemOpen
  \bibfield  {author} {\bibinfo {author} {\bibfnamefont {H.}~\bibnamefont
  {Nishimori}}, \bibinfo {author} {\bibfnamefont {K.}~\bibnamefont {Uchiyama}},
  \bibinfo {author} {\bibfnamefont {S.}~\bibnamefont {Kaneko}}, \bibinfo
  {author} {\bibfnamefont {A.}~\bibnamefont {Tokura}}, \bibinfo {author}
  {\bibfnamefont {H.}~\bibnamefont {Takeya}}, \bibinfo {author} {\bibfnamefont
  {K.}~\bibnamefont {Hirata}},\ and\ \bibinfo {author} {\bibfnamefont
  {N.}~\bibnamefont {Nishida}},\ }\href@noop {} {\bibfield  {journal} {\bibinfo
   {journal} {Journal of the Physical Society of Japan}\ }\textbf {\bibinfo
  {volume} {73}},\ \bibinfo {pages} {3247} (\bibinfo {year}
  {2004})}\BibitemShut {NoStop}%
\bibitem [{\citenamefont {De~Wilde}\ \emph {et~al.}(1997)\citenamefont
  {De~Wilde}, \citenamefont {Iavarone}, \citenamefont {Welp}, \citenamefont
  {Metlushko}, \citenamefont {Koshelev}, \citenamefont {Aranson}, \citenamefont
  {Crabtree},\ and\ \citenamefont {Canfield}}]{PhysRevLett.78.4273}%
  \BibitemOpen
  \bibfield  {author} {\bibinfo {author} {\bibfnamefont {Y.}~\bibnamefont
  {De~Wilde}}, \bibinfo {author} {\bibfnamefont {M.}~\bibnamefont {Iavarone}},
  \bibinfo {author} {\bibfnamefont {U.}~\bibnamefont {Welp}}, \bibinfo {author}
  {\bibfnamefont {V.}~\bibnamefont {Metlushko}}, \bibinfo {author}
  {\bibfnamefont {A.~E.}\ \bibnamefont {Koshelev}}, \bibinfo {author}
  {\bibfnamefont {I.}~\bibnamefont {Aranson}}, \bibinfo {author} {\bibfnamefont
  {G.~W.}\ \bibnamefont {Crabtree}},\ and\ \bibinfo {author} {\bibfnamefont
  {P.~C.}\ \bibnamefont {Canfield}},\ }\href@noop {} {\bibfield  {journal}
  {\bibinfo  {journal} {Phys. Rev. Lett.}\ }\textbf {\bibinfo {volume} {78}},\
  \bibinfo {pages} {4273} (\bibinfo {year} {1997})}\BibitemShut {NoStop}%
\bibitem [{\citenamefont {Suderow}\ \emph {et~al.}(2014)\citenamefont
  {Suderow}, \citenamefont {Guillam{\'o}n}, \citenamefont {Rodrigo},\ and\
  \citenamefont {Vieira}}]{suderow2014imaging}%
  \BibitemOpen
  \bibfield  {author} {\bibinfo {author} {\bibfnamefont {H.}~\bibnamefont
  {Suderow}}, \bibinfo {author} {\bibfnamefont {I.}~\bibnamefont
  {Guillam{\'o}n}}, \bibinfo {author} {\bibfnamefont {J.~G.}\ \bibnamefont
  {Rodrigo}},\ and\ \bibinfo {author} {\bibfnamefont {S.}~\bibnamefont
  {Vieira}},\ }\href@noop {} {\bibfield  {journal} {\bibinfo  {journal}
  {Superconductor Science and Technology}\ }\textbf {\bibinfo {volume} {27}},\
  \bibinfo {pages} {063001} (\bibinfo {year} {2014})}\BibitemShut {NoStop}%
\bibitem [{\citenamefont {Maggio-Aprile}\ \emph {et~al.}(1995)\citenamefont
  {Maggio-Aprile}, \citenamefont {Renner}, \citenamefont {Erb}, \citenamefont
  {Walker},\ and\ \citenamefont {Fischer}}]{PhysRevLett.75.2754}%
  \BibitemOpen
  \bibfield  {author} {\bibinfo {author} {\bibfnamefont {I.}~\bibnamefont
  {Maggio-Aprile}}, \bibinfo {author} {\bibfnamefont {C.}~\bibnamefont
  {Renner}}, \bibinfo {author} {\bibfnamefont {A.}~\bibnamefont {Erb}},
  \bibinfo {author} {\bibfnamefont {E.}~\bibnamefont {Walker}},\ and\ \bibinfo
  {author} {\bibfnamefont {O.}~\bibnamefont {Fischer}},\ }\href@noop {}
  {\bibfield  {journal} {\bibinfo  {journal} {Phys. Rev. Lett.}\ }\textbf
  {\bibinfo {volume} {75}},\ \bibinfo {pages} {2754} (\bibinfo {year}
  {1995})}\BibitemShut {NoStop}%
\bibitem [{\citenamefont {Berthod}\ and\ \citenamefont
  {Giovannini}(2001)}]{PhysRevLett.87.277002}%
  \BibitemOpen
  \bibfield  {author} {\bibinfo {author} {\bibfnamefont {C.}~\bibnamefont
  {Berthod}}\ and\ \bibinfo {author} {\bibfnamefont {B.}~\bibnamefont
  {Giovannini}},\ }\href@noop {} {\bibfield  {journal} {\bibinfo  {journal}
  {Phys. Rev. Lett.}\ }\textbf {\bibinfo {volume} {87}},\ \bibinfo {pages}
  {277002} (\bibinfo {year} {2001})}\BibitemShut {NoStop}%
\bibitem [{\citenamefont {Fischer}\ \emph {et~al.}(2007)\citenamefont
  {Fischer}, \citenamefont {Kugler}, \citenamefont {Maggio-Aprile},
  \citenamefont {Berthod},\ and\ \citenamefont {Renner}}]{RevModPhys.79.353}%
  \BibitemOpen
  \bibfield  {author} {\bibinfo {author} {\bibfnamefont {O.}~\bibnamefont
  {Fischer}}, \bibinfo {author} {\bibfnamefont {M.}~\bibnamefont {Kugler}},
  \bibinfo {author} {\bibfnamefont {I.}~\bibnamefont {Maggio-Aprile}}, \bibinfo
  {author} {\bibfnamefont {C.}~\bibnamefont {Berthod}},\ and\ \bibinfo {author}
  {\bibfnamefont {C.}~\bibnamefont {Renner}},\ }\href@noop {} {\bibfield
  {journal} {\bibinfo  {journal} {Rev. Mod. Phys.}\ }\textbf {\bibinfo {volume}
  {79}},\ \bibinfo {pages} {353} (\bibinfo {year} {2007})}\BibitemShut
  {NoStop}%
\bibitem [{\citenamefont {Putilov}\ \emph {et~al.}(2019)\citenamefont
  {Putilov}, \citenamefont {Di~Giorgio}, \citenamefont {Vadimov}, \citenamefont
  {Trainer}, \citenamefont {Lechner}, \citenamefont {Curtis}, \citenamefont
  {Abdel-Hafiez}, \citenamefont {Volkova}, \citenamefont {Vasiliev},
  \citenamefont {Chareev}, \citenamefont {Karapetrov}, \citenamefont
  {Koshelev}, \citenamefont {Aladyshkin}, \citenamefont {Mel'nikov},\ and\
  \citenamefont {Iavarone}}]{PhysRevB.99.144514}%
  \BibitemOpen
  \bibfield  {author} {\bibinfo {author} {\bibfnamefont {A.~V.}\ \bibnamefont
  {Putilov}}, \bibinfo {author} {\bibfnamefont {C.}~\bibnamefont {Di~Giorgio}},
  \bibinfo {author} {\bibfnamefont {V.~L.}\ \bibnamefont {Vadimov}}, \bibinfo
  {author} {\bibfnamefont {D.~J.}\ \bibnamefont {Trainer}}, \bibinfo {author}
  {\bibfnamefont {E.~M.}\ \bibnamefont {Lechner}}, \bibinfo {author}
  {\bibfnamefont {J.~L.}\ \bibnamefont {Curtis}}, \bibinfo {author}
  {\bibfnamefont {M.}~\bibnamefont {Abdel-Hafiez}}, \bibinfo {author}
  {\bibfnamefont {O.~S.}\ \bibnamefont {Volkova}}, \bibinfo {author}
  {\bibfnamefont {A.~N.}\ \bibnamefont {Vasiliev}}, \bibinfo {author}
  {\bibfnamefont {D.~A.}\ \bibnamefont {Chareev}}, \bibinfo {author}
  {\bibfnamefont {G.}~\bibnamefont {Karapetrov}}, \bibinfo {author}
  {\bibfnamefont {A.~E.}\ \bibnamefont {Koshelev}}, \bibinfo {author}
  {\bibfnamefont {A.~Y.}\ \bibnamefont {Aladyshkin}}, \bibinfo {author}
  {\bibfnamefont {A.~S.}\ \bibnamefont {Mel'nikov}},\ and\ \bibinfo {author}
  {\bibfnamefont {M.}~\bibnamefont {Iavarone}},\ }\href@noop {} {\bibfield
  {journal} {\bibinfo  {journal} {Phys. Rev. B}\ }\textbf {\bibinfo {volume}
  {99}},\ \bibinfo {pages} {144514} (\bibinfo {year} {2019})}\BibitemShut
  {NoStop}%
\bibitem [{\citenamefont {Berthod}\ \emph {et~al.}(2017)\citenamefont
  {Berthod}, \citenamefont {Maggio-Aprile}, \citenamefont {Bru\'er},
  \citenamefont {Erb},\ and\ \citenamefont {Renner}}]{PhysRevLett.119.237001}%
  \BibitemOpen
  \bibfield  {author} {\bibinfo {author} {\bibfnamefont {C.}~\bibnamefont
  {Berthod}}, \bibinfo {author} {\bibfnamefont {I.}~\bibnamefont
  {Maggio-Aprile}}, \bibinfo {author} {\bibfnamefont {J.}~\bibnamefont
  {Bru\'er}}, \bibinfo {author} {\bibfnamefont {A.}~\bibnamefont {Erb}},\ and\
  \bibinfo {author} {\bibfnamefont {C.}~\bibnamefont {Renner}},\ }\href@noop {}
  {\bibfield  {journal} {\bibinfo  {journal} {Phys. Rev. Lett.}\ }\textbf
  {\bibinfo {volume} {119}},\ \bibinfo {pages} {237001} (\bibinfo {year}
  {2017})}\BibitemShut {NoStop}%
\bibitem [{\citenamefont {Volovik}(2009)}]{volovik2009universe}%
  \BibitemOpen
  \bibfield  {author} {\bibinfo {author} {\bibfnamefont {G.}~\bibnamefont
  {Volovik}},\ }\href@noop {} {\emph {\bibinfo {title} {The Universe in a
  Helium Droplet}}},\ International Series of Monographs on Physics\ (\bibinfo
  {publisher} {OUP Oxford},\ \bibinfo {year} {2009})\BibitemShut {NoStop}%
\bibitem [{\citenamefont {Blatter}\ \emph {et~al.}(1994)\citenamefont
  {Blatter}, \citenamefont {Feigel'man}, \citenamefont {Geshkenbein},
  \citenamefont {Larkin},\ and\ \citenamefont {Vinokur}}]{RevModPhys.66.1125}%
  \BibitemOpen
  \bibfield  {author} {\bibinfo {author} {\bibfnamefont {G.}~\bibnamefont
  {Blatter}}, \bibinfo {author} {\bibfnamefont {M.~V.}\ \bibnamefont
  {Feigel'man}}, \bibinfo {author} {\bibfnamefont {V.~B.}\ \bibnamefont
  {Geshkenbein}}, \bibinfo {author} {\bibfnamefont {A.~I.}\ \bibnamefont
  {Larkin}},\ and\ \bibinfo {author} {\bibfnamefont {V.~M.}\ \bibnamefont
  {Vinokur}},\ }\href@noop {} {\bibfield  {journal} {\bibinfo  {journal} {Rev.
  Mod. Phys.}\ }\textbf {\bibinfo {volume} {66}},\ \bibinfo {pages} {1125}
  (\bibinfo {year} {1994})}\BibitemShut {NoStop}%
\bibitem [{\citenamefont {Brandt}(1995)}]{Brandt_1995}%
  \BibitemOpen
  \bibfield  {author} {\bibinfo {author} {\bibfnamefont {E.~H.}\ \bibnamefont
  {Brandt}},\ }\href@noop {} {\bibfield  {journal} {\bibinfo  {journal}
  {Reports on Progress in Physics}\ }\textbf {\bibinfo {volume} {58}},\
  \bibinfo {pages} {1465} (\bibinfo {year} {1995})}\BibitemShut {NoStop}%
\bibitem [{\citenamefont {Roditchev}\ \emph {et~al.}(2015)\citenamefont
  {Roditchev}, \citenamefont {Brun}, \citenamefont {Serrier-Garcia},
  \citenamefont {Cuevas}, \citenamefont {Bessa}, \citenamefont
  {Milo{\v{s}}evi{\'{c}}}, \citenamefont {Debontridder}, \citenamefont
  {Stolyarov},\ and\ \citenamefont {Cren}}]{Roditchev2015}%
  \BibitemOpen
  \bibfield  {author} {\bibinfo {author} {\bibfnamefont {D.}~\bibnamefont
  {Roditchev}}, \bibinfo {author} {\bibfnamefont {C.}~\bibnamefont {Brun}},
  \bibinfo {author} {\bibfnamefont {L.}~\bibnamefont {Serrier-Garcia}},
  \bibinfo {author} {\bibfnamefont {J.~C.}\ \bibnamefont {Cuevas}}, \bibinfo
  {author} {\bibfnamefont {V.~H.~L.}\ \bibnamefont {Bessa}}, \bibinfo {author}
  {\bibfnamefont {M.~V.}\ \bibnamefont {Milo{\v{s}}evi{\'{c}}}}, \bibinfo
  {author} {\bibfnamefont {F.}~\bibnamefont {Debontridder}}, \bibinfo {author}
  {\bibfnamefont {V.}~\bibnamefont {Stolyarov}},\ and\ \bibinfo {author}
  {\bibfnamefont {T.}~\bibnamefont {Cren}},\ }\href@noop {} {\bibfield
  {journal} {\bibinfo  {journal} {Nature Physics}\ }\textbf {\bibinfo {volume}
  {11}},\ \bibinfo {pages} {332} (\bibinfo {year} {2015})}\BibitemShut
  {NoStop}%
\bibitem [{\citenamefont {Kramer}\ and\ \citenamefont
  {Pesch}(1974)}]{Kramer1974}%
  \BibitemOpen
  \bibfield  {author} {\bibinfo {author} {\bibfnamefont {L.}~\bibnamefont
  {Kramer}}\ and\ \bibinfo {author} {\bibfnamefont {W.}~\bibnamefont {Pesch}},\
  }\href@noop {} {\bibfield  {journal} {\bibinfo  {journal} {Zeitschrift
  f{\"u}r Physik}\ }\textbf {\bibinfo {volume} {269}},\ \bibinfo {pages} {59}
  (\bibinfo {year} {1974})}\BibitemShut {NoStop}%
\bibitem [{\citenamefont {Klein}(1990)}]{PhysRevB.41.4819}%
  \BibitemOpen
  \bibfield  {author} {\bibinfo {author} {\bibfnamefont {U.}~\bibnamefont
  {Klein}},\ }\href@noop {} {\bibfield  {journal} {\bibinfo  {journal} {Phys.
  Rev. B}\ }\textbf {\bibinfo {volume} {41}},\ \bibinfo {pages} {4819}
  (\bibinfo {year} {1990})}\BibitemShut {NoStop}%
\bibitem [{\citenamefont {Hess}\ \emph {et~al.}(1990)\citenamefont {Hess},
  \citenamefont {Robinson},\ and\ \citenamefont
  {Waszczak}}]{PhysRevLett.64.2711}%
  \BibitemOpen
  \bibfield  {author} {\bibinfo {author} {\bibfnamefont {H.~F.}\ \bibnamefont
  {Hess}}, \bibinfo {author} {\bibfnamefont {R.~B.}\ \bibnamefont {Robinson}},\
  and\ \bibinfo {author} {\bibfnamefont {J.~V.}\ \bibnamefont {Waszczak}},\
  }\href@noop {} {\bibfield  {journal} {\bibinfo  {journal} {Phys. Rev. Lett.}\
  }\textbf {\bibinfo {volume} {64}},\ \bibinfo {pages} {2711} (\bibinfo {year}
  {1990})}\BibitemShut {NoStop}%
\bibitem [{\citenamefont {Caroli}\ \emph {et~al.}(1964)\citenamefont {Caroli},
  \citenamefont {{De Gennes}},\ and\ \citenamefont {Matricon}}]{CAROLI1964307}%
  \BibitemOpen
  \bibfield  {author} {\bibinfo {author} {\bibfnamefont {C.}~\bibnamefont
  {Caroli}}, \bibinfo {author} {\bibfnamefont {P.}~\bibnamefont {{De
  Gennes}}},\ and\ \bibinfo {author} {\bibfnamefont {J.}~\bibnamefont
  {Matricon}},\ }\href@noop {} {\bibfield  {journal} {\bibinfo  {journal}
  {Physics Letters}\ }\textbf {\bibinfo {volume} {9}},\ \bibinfo {pages} {307}
  (\bibinfo {year} {1964})}\BibitemShut {NoStop}%
\bibitem [{\citenamefont {Larkin}\ and\ \citenamefont
  {Ovchinnikov}(1998)}]{PhysRevB.57.5457}%
  \BibitemOpen
  \bibfield  {author} {\bibinfo {author} {\bibfnamefont {A.~I.}\ \bibnamefont
  {Larkin}}\ and\ \bibinfo {author} {\bibfnamefont {Y.~N.}\ \bibnamefont
  {Ovchinnikov}},\ }\href@noop {} {\bibfield  {journal} {\bibinfo  {journal}
  {Phys. Rev. B}\ }\textbf {\bibinfo {volume} {57}},\ \bibinfo {pages} {5457}
  (\bibinfo {year} {1998})}\BibitemShut {NoStop}%
\bibitem [{\citenamefont {Skvortsov}\ \emph {et~al.}(1998)\citenamefont
  {Skvortsov}, \citenamefont {Feigel'man},\ and\ \citenamefont
  {Kravtsov}}]{Skvortsov1998}%
  \BibitemOpen
  \bibfield  {author} {\bibinfo {author} {\bibfnamefont {M.~A.}\ \bibnamefont
  {Skvortsov}}, \bibinfo {author} {\bibfnamefont {M.~V.}\ \bibnamefont
  {Feigel'man}},\ and\ \bibinfo {author} {\bibfnamefont {V.~E.}\ \bibnamefont
  {Kravtsov}},\ }\href@noop {} {\bibfield  {journal} {\bibinfo  {journal} {JETP
  Lett.}\ }\textbf {\bibinfo {volume} {68}},\ \bibinfo {pages} {84} (\bibinfo
  {year} {1998})}\BibitemShut {NoStop}%
\bibitem [{\citenamefont {Koulakov}\ and\ \citenamefont
  {Larkin}(1999)}]{PhysRevB.60.14597}%
  \BibitemOpen
  \bibfield  {author} {\bibinfo {author} {\bibfnamefont {A.~A.}\ \bibnamefont
  {Koulakov}}\ and\ \bibinfo {author} {\bibfnamefont {A.~I.}\ \bibnamefont
  {Larkin}},\ }\href@noop {} {\bibfield  {journal} {\bibinfo  {journal} {Phys.
  Rev. B}\ }\textbf {\bibinfo {volume} {60}},\ \bibinfo {pages} {14597}
  (\bibinfo {year} {1999})}\BibitemShut {NoStop}%
\bibitem [{\citenamefont {Bespalov}\ and\ \citenamefont
  {Plastovets}(2021)}]{PhysRevB.103.024510}%
  \BibitemOpen
  \bibfield  {author} {\bibinfo {author} {\bibfnamefont {A.~A.}\ \bibnamefont
  {Bespalov}}\ and\ \bibinfo {author} {\bibfnamefont {V.~D.}\ \bibnamefont
  {Plastovets}},\ }\href@noop {} {\bibfield  {journal} {\bibinfo  {journal}
  {Phys. Rev. B}\ }\textbf {\bibinfo {volume} {103}},\ \bibinfo {pages}
  {024510} (\bibinfo {year} {2021})}\BibitemShut {NoStop}%
\bibitem [{\citenamefont {Mel'nikov}\ \emph {et~al.}(2009)\citenamefont
  {Mel'nikov}, \citenamefont {Samokhvalov},\ and\ \citenamefont
  {Zubarev}}]{PhysRevB.79.134529}%
  \BibitemOpen
  \bibfield  {author} {\bibinfo {author} {\bibfnamefont {A.~S.}\ \bibnamefont
  {Mel'nikov}}, \bibinfo {author} {\bibfnamefont {A.~V.}\ \bibnamefont
  {Samokhvalov}},\ and\ \bibinfo {author} {\bibfnamefont {M.~N.}\ \bibnamefont
  {Zubarev}},\ }\href@noop {} {\bibfield  {journal} {\bibinfo  {journal} {Phys.
  Rev. B}\ }\textbf {\bibinfo {volume} {79}},\ \bibinfo {pages} {134529}
  (\bibinfo {year} {2009})}\BibitemShut {NoStop}%
\bibitem [{\citenamefont {Rosenstein}\ \emph {et~al.}(2011)\citenamefont
  {Rosenstein}, \citenamefont {Shapiro}, \citenamefont {Deutch},\ and\
  \citenamefont {Shapiro}}]{PhysRevB.84.134521}%
  \BibitemOpen
  \bibfield  {author} {\bibinfo {author} {\bibfnamefont {B.}~\bibnamefont
  {Rosenstein}}, \bibinfo {author} {\bibfnamefont {I.}~\bibnamefont {Shapiro}},
  \bibinfo {author} {\bibfnamefont {E.}~\bibnamefont {Deutch}},\ and\ \bibinfo
  {author} {\bibfnamefont {B.~Y.}\ \bibnamefont {Shapiro}},\ }\href@noop {}
  {\bibfield  {journal} {\bibinfo  {journal} {Phys. Rev. B}\ }\textbf {\bibinfo
  {volume} {84}},\ \bibinfo {pages} {134521} (\bibinfo {year}
  {2011})}\BibitemShut {NoStop}%
\bibitem [{\citenamefont {Samokhvalov}\ \emph {et~al.}(2020)\citenamefont
  {Samokhvalov}, \citenamefont {Plastovets},\ and\ \citenamefont
  {Mel'nikov}}]{PhysRevB.102.174501}%
  \BibitemOpen
  \bibfield  {author} {\bibinfo {author} {\bibfnamefont {A.~V.}\ \bibnamefont
  {Samokhvalov}}, \bibinfo {author} {\bibfnamefont {V.~D.}\ \bibnamefont
  {Plastovets}},\ and\ \bibinfo {author} {\bibfnamefont {A.~S.}\ \bibnamefont
  {Mel'nikov}},\ }\href@noop {} {\bibfield  {journal} {\bibinfo  {journal}
  {Phys. Rev. B}\ }\textbf {\bibinfo {volume} {102}},\ \bibinfo {pages}
  {174501} (\bibinfo {year} {2020})}\BibitemShut {NoStop}%
\bibitem [{\citenamefont {Khodaeva}\ and\ \citenamefont
  {Skvortsov}()}]{khodaeva2021vortex}%
  \BibitemOpen
  \bibfield  {author} {\bibinfo {author} {\bibfnamefont {U.~E.}\ \bibnamefont
  {Khodaeva}}\ and\ \bibinfo {author} {\bibfnamefont {M.~A.}\ \bibnamefont
  {Skvortsov}},\ }\href@noop {} {}\Eprint {https://arxiv.org/abs/2112.06303
  (2021)} {arXiv:2112.06303 (2021)} \BibitemShut {NoStop}%
\bibitem [{\citenamefont {Graser}\ \emph {et~al.}(2004)\citenamefont {Graser},
  \citenamefont {Iniotakis}, \citenamefont {Dahm},\ and\ \citenamefont
  {Schopohl}}]{PhysRevLett.93.247001}%
  \BibitemOpen
  \bibfield  {author} {\bibinfo {author} {\bibfnamefont {S.}~\bibnamefont
  {Graser}}, \bibinfo {author} {\bibfnamefont {C.}~\bibnamefont {Iniotakis}},
  \bibinfo {author} {\bibfnamefont {T.}~\bibnamefont {Dahm}},\ and\ \bibinfo
  {author} {\bibfnamefont {N.}~\bibnamefont {Schopohl}},\ }\href@noop {}
  {\bibfield  {journal} {\bibinfo  {journal} {Phys. Rev. Lett.}\ }\textbf
  {\bibinfo {volume} {93}},\ \bibinfo {pages} {247001} (\bibinfo {year}
  {2004})}\BibitemShut {NoStop}%
\bibitem [{\citenamefont {Mel'nikov}\ \emph {et~al.}(2008)\citenamefont
  {Mel'nikov}, \citenamefont {Ryzhov},\ and\ \citenamefont
  {Silaev}}]{PhysRevB.78.064513}%
  \BibitemOpen
  \bibfield  {author} {\bibinfo {author} {\bibfnamefont {A.~S.}\ \bibnamefont
  {Mel'nikov}}, \bibinfo {author} {\bibfnamefont {D.~A.}\ \bibnamefont
  {Ryzhov}},\ and\ \bibinfo {author} {\bibfnamefont {M.~A.}\ \bibnamefont
  {Silaev}},\ }\href@noop {} {\bibfield  {journal} {\bibinfo  {journal} {Phys.
  Rev. B}\ }\textbf {\bibinfo {volume} {78}},\ \bibinfo {pages} {064513}
  (\bibinfo {year} {2008})}\BibitemShut {NoStop}%
\bibitem [{\citenamefont {Iniotakis}\ \emph {et~al.}(2005)\citenamefont
  {Iniotakis}, \citenamefont {Graser}, \citenamefont {Dahm},\ and\
  \citenamefont {Schopohl}}]{PhysRevB.71.214508}%
  \BibitemOpen
  \bibfield  {author} {\bibinfo {author} {\bibfnamefont {C.}~\bibnamefont
  {Iniotakis}}, \bibinfo {author} {\bibfnamefont {S.}~\bibnamefont {Graser}},
  \bibinfo {author} {\bibfnamefont {T.}~\bibnamefont {Dahm}},\ and\ \bibinfo
  {author} {\bibfnamefont {N.}~\bibnamefont {Schopohl}},\ }\href@noop {}
  {\bibfield  {journal} {\bibinfo  {journal} {Phys. Rev. B}\ }\textbf {\bibinfo
  {volume} {71}},\ \bibinfo {pages} {214508} (\bibinfo {year}
  {2005})}\BibitemShut {NoStop}%
\bibitem [{\citenamefont {Rakhmanov}\ \emph {et~al.}(2011)\citenamefont
  {Rakhmanov}, \citenamefont {Rozhkov},\ and\ \citenamefont
  {Nori}}]{PhysRevB.84.075141}%
  \BibitemOpen
  \bibfield  {author} {\bibinfo {author} {\bibfnamefont {A.~L.}\ \bibnamefont
  {Rakhmanov}}, \bibinfo {author} {\bibfnamefont {A.~V.}\ \bibnamefont
  {Rozhkov}},\ and\ \bibinfo {author} {\bibfnamefont {F.}~\bibnamefont
  {Nori}},\ }\href@noop {} {\bibfield  {journal} {\bibinfo  {journal} {Phys.
  Rev. B}\ }\textbf {\bibinfo {volume} {84}},\ \bibinfo {pages} {075141}
  (\bibinfo {year} {2011})}\BibitemShut {NoStop}%
\bibitem [{\citenamefont {Ioselevich}\ and\ \citenamefont
  {Feigel'man}(2011)}]{PhysRevLett.106.077003}%
  \BibitemOpen
  \bibfield  {author} {\bibinfo {author} {\bibfnamefont {P.~A.}\ \bibnamefont
  {Ioselevich}}\ and\ \bibinfo {author} {\bibfnamefont {M.~V.}\ \bibnamefont
  {Feigel'man}},\ }\href@noop {} {\bibfield  {journal} {\bibinfo  {journal}
  {Phys. Rev. Lett.}\ }\textbf {\bibinfo {volume} {106}},\ \bibinfo {pages}
  {077003} (\bibinfo {year} {2011})}\BibitemShut {NoStop}%
\bibitem [{\citenamefont {Ioselevich}\ \emph {et~al.}(2012)\citenamefont
  {Ioselevich}, \citenamefont {Ostrovsky},\ and\ \citenamefont
  {Feigel'man}}]{PhysRevB.86.035441}%
  \BibitemOpen
  \bibfield  {author} {\bibinfo {author} {\bibfnamefont {P.~A.}\ \bibnamefont
  {Ioselevich}}, \bibinfo {author} {\bibfnamefont {P.~M.}\ \bibnamefont
  {Ostrovsky}},\ and\ \bibinfo {author} {\bibfnamefont {M.~V.}\ \bibnamefont
  {Feigel'man}},\ }\href@noop {} {\bibfield  {journal} {\bibinfo  {journal}
  {Phys. Rev. B}\ }\textbf {\bibinfo {volume} {86}},\ \bibinfo {pages} {035441}
  (\bibinfo {year} {2012})}\BibitemShut {NoStop}%
\bibitem [{\citenamefont {Akzyanov}\ \emph {et~al.}(2014)\citenamefont
  {Akzyanov}, \citenamefont {Rozhkov}, \citenamefont {Rakhmanov},\ and\
  \citenamefont {Nori}}]{PhysRevB.89.085409}%
  \BibitemOpen
  \bibfield  {author} {\bibinfo {author} {\bibfnamefont {R.~S.}\ \bibnamefont
  {Akzyanov}}, \bibinfo {author} {\bibfnamefont {A.~V.}\ \bibnamefont
  {Rozhkov}}, \bibinfo {author} {\bibfnamefont {A.~L.}\ \bibnamefont
  {Rakhmanov}},\ and\ \bibinfo {author} {\bibfnamefont {F.}~\bibnamefont
  {Nori}},\ }\href@noop {} {\bibfield  {journal} {\bibinfo  {journal} {Phys.
  Rev. B}\ }\textbf {\bibinfo {volume} {89}},\ \bibinfo {pages} {085409}
  (\bibinfo {year} {2014})}\BibitemShut {NoStop}%
\bibitem [{\citenamefont {Fu}\ and\ \citenamefont
  {Kane}(2008)}]{PhysRevLett.100.096407}%
  \BibitemOpen
  \bibfield  {author} {\bibinfo {author} {\bibfnamefont {L.}~\bibnamefont
  {Fu}}\ and\ \bibinfo {author} {\bibfnamefont {C.~L.}\ \bibnamefont {Kane}},\
  }\href@noop {} {\bibfield  {journal} {\bibinfo  {journal} {Phys. Rev. Lett.}\
  }\textbf {\bibinfo {volume} {100}},\ \bibinfo {pages} {096407} (\bibinfo
  {year} {2008})}\BibitemShut {NoStop}%
\bibitem [{\citenamefont {Sau}\ \emph {et~al.}(2010)\citenamefont {Sau},
  \citenamefont {Lutchyn}, \citenamefont {Tewari},\ and\ \citenamefont
  {Das~Sarma}}]{PhysRevB.82.094522}%
  \BibitemOpen
  \bibfield  {author} {\bibinfo {author} {\bibfnamefont {J.~D.}\ \bibnamefont
  {Sau}}, \bibinfo {author} {\bibfnamefont {R.~M.}\ \bibnamefont {Lutchyn}},
  \bibinfo {author} {\bibfnamefont {S.}~\bibnamefont {Tewari}},\ and\ \bibinfo
  {author} {\bibfnamefont {S.}~\bibnamefont {Das~Sarma}},\ }\href@noop {}
  {\bibfield  {journal} {\bibinfo  {journal} {Phys. Rev. B}\ }\textbf {\bibinfo
  {volume} {82}},\ \bibinfo {pages} {094522} (\bibinfo {year}
  {2010})}\BibitemShut {NoStop}%
\bibitem [{\citenamefont {Gurevich}(1992)}]{PhysRevB.46.3187}%
  \BibitemOpen
  \bibfield  {author} {\bibinfo {author} {\bibfnamefont {A.}~\bibnamefont
  {Gurevich}},\ }\href@noop {} {\bibfield  {journal} {\bibinfo  {journal}
  {Phys. Rev. B}\ }\textbf {\bibinfo {volume} {46}},\ \bibinfo {pages} {3187}
  (\bibinfo {year} {1992})}\BibitemShut {NoStop}%
\bibitem [{\citenamefont {Horide}\ \emph {et~al.}(2007)\citenamefont {Horide},
  \citenamefont {Matsumoto}, \citenamefont {Ichinose}, \citenamefont {Mukaida},
  \citenamefont {Yoshida},\ and\ \citenamefont {Horii}}]{PhysRevB.75.020504}%
  \BibitemOpen
  \bibfield  {author} {\bibinfo {author} {\bibfnamefont {T.}~\bibnamefont
  {Horide}}, \bibinfo {author} {\bibfnamefont {K.}~\bibnamefont {Matsumoto}},
  \bibinfo {author} {\bibfnamefont {A.}~\bibnamefont {Ichinose}}, \bibinfo
  {author} {\bibfnamefont {M.}~\bibnamefont {Mukaida}}, \bibinfo {author}
  {\bibfnamefont {Y.}~\bibnamefont {Yoshida}},\ and\ \bibinfo {author}
  {\bibfnamefont {S.}~\bibnamefont {Horii}},\ }\href@noop {} {\bibfield
  {journal} {\bibinfo  {journal} {Phys. Rev. B}\ }\textbf {\bibinfo {volume}
  {75}},\ \bibinfo {pages} {020504} (\bibinfo {year} {2007})}\BibitemShut
  {NoStop}%
\bibitem [{\citenamefont {Horide}\ \emph {et~al.}(2008)\citenamefont {Horide},
  \citenamefont {Matsumoto}, \citenamefont {Yoshida}, \citenamefont {Mukaida},
  \citenamefont {Ichinose},\ and\ \citenamefont {Horii}}]{PhysRevB.77.132502}%
  \BibitemOpen
  \bibfield  {author} {\bibinfo {author} {\bibfnamefont {T.}~\bibnamefont
  {Horide}}, \bibinfo {author} {\bibfnamefont {K.}~\bibnamefont {Matsumoto}},
  \bibinfo {author} {\bibfnamefont {Y.}~\bibnamefont {Yoshida}}, \bibinfo
  {author} {\bibfnamefont {M.}~\bibnamefont {Mukaida}}, \bibinfo {author}
  {\bibfnamefont {A.}~\bibnamefont {Ichinose}},\ and\ \bibinfo {author}
  {\bibfnamefont {S.}~\bibnamefont {Horii}},\ }\href@noop {} {\bibfield
  {journal} {\bibinfo  {journal} {Phys. Rev. B}\ }\textbf {\bibinfo {volume}
  {77}},\ \bibinfo {pages} {132502} (\bibinfo {year} {2008})}\BibitemShut
  {NoStop}%
\bibitem [{\citenamefont {Beenakker}\ and\ \citenamefont {van
  Houten}(1991)}]{PhysRevLett.66.3056}%
  \BibitemOpen
  \bibfield  {author} {\bibinfo {author} {\bibfnamefont {C.~W.~J.}\
  \bibnamefont {Beenakker}}\ and\ \bibinfo {author} {\bibfnamefont
  {H.}~\bibnamefont {van Houten}},\ }\href@noop {} {\bibfield  {journal}
  {\bibinfo  {journal} {Phys. Rev. Lett.}\ }\textbf {\bibinfo {volume} {66}},\
  \bibinfo {pages} {3056} (\bibinfo {year} {1991})}\BibitemShut {NoStop}%
\bibitem [{\citenamefont {Beenakker}(1991)}]{PhysRevLett.67.3836}%
  \BibitemOpen
  \bibfield  {author} {\bibinfo {author} {\bibfnamefont {C.~W.~J.}\
  \bibnamefont {Beenakker}},\ }\href@noop {} {\bibfield  {journal} {\bibinfo
  {journal} {Phys. Rev. Lett.}\ }\textbf {\bibinfo {volume} {67}},\ \bibinfo
  {pages} {3836} (\bibinfo {year} {1991})}\BibitemShut {NoStop}%
\bibitem [{\citenamefont {Haberman}(2012)}]{Fredholm}%
  \BibitemOpen
  \bibfield  {author} {\bibinfo {author} {\bibfnamefont {R.}~\bibnamefont
  {Haberman}},\ }\href@noop {} {\emph {\bibinfo {title} {Applied partial
  differential equations with Fourier Series and Boundary Value Problems (5th
  edition)}}}\ (\bibinfo  {publisher} {Pearson},\ \bibinfo {year}
  {2012})\BibitemShut {NoStop}%
\bibitem [{\citenamefont {Landau}\ and\ \citenamefont
  {Lifshitz}(1991)}]{Landau}%
  \BibitemOpen
  \bibfield  {author} {\bibinfo {author} {\bibfnamefont {L.~D.}\ \bibnamefont
  {Landau}}\ and\ \bibinfo {author} {\bibfnamefont {E.~M.}\ \bibnamefont
  {Lifshitz}},\ }\href@noop {} {\emph {\bibinfo {title} {Quantum Mechanics.
  Nonrelativistic theory}}}\ (\bibinfo  {publisher} {Pergamon Press},\ \bibinfo
  {year} {1991})\BibitemShut {NoStop}%
\bibitem [{\citenamefont {Tinkham}(2004)}]{Tinkham04}%
  \BibitemOpen
  \bibfield  {author} {\bibinfo {author} {\bibfnamefont {M.}~\bibnamefont
  {Tinkham}},\ }\href@noop {} {\emph {\bibinfo {title} {Introduction to
  Superconductivity (2nd edition)}}}\ (\bibinfo  {publisher} {Dover
  Publication},\ \bibinfo {year} {2004})\BibitemShut {NoStop}%
\end{thebibliography}

\end{document}